\documentclass[number,11pt,1p]{elsarticle}

\usepackage{graphics}
\usepackage{lineno} 
\usepackage{ulem}
\usepackage{tabularx}
\usepackage{textcomp}
\usepackage{amssymb}
\usepackage{units}

\usepackage{subfig}
\usepackage[ps2pdf,%
colorlinks=true,%
pdfauthor={Daniel Huber},%
pdftitle={NIMA}%
]{hyperref}

\begin{document}

\begin{frontmatter}

\title{LOPES-3D, an antenna array for full signal detection of air-shower radio 
emission}

\author[1]{W.D.~Apel}
\author[2,14]{J.C.~Arteaga}
\author[3]{L.~B\"ahren}
\author[1]{K.~Bekk}
\author[4]{M.~Bertaina}
\author[5]{P.L.~Biermann}
\author[1,2]{J.~Bl\"umer}
\author[1]{H.~Bozdog}
\author[6]{I.M.~Brancus}
\author[7]{P.~Buchholz}
\author[4,8]{E.~Cantoni}
\author[4]{A.~Chiavassa}
\author[1]{K.~Daumiller}
\author[2,15]{V.~de~Souza}
\author[4]{F.~Di~Pierro}
\author[1]{P.~Doll}
\author[1]{R.~Engel}
\author[3,9,5]{H.~Falcke}
\author[2]{M. Finger}
\author[2]{B.~Fuchs}
\author[10]{D.~Fuhrmann}
\author[11]{H.~Gemmeke}
\author[7]{C.~Grupen}
\author[1]{A.~Haungs}
\author[1]{D.~Heck}
\author[3]{J.R.~H\"orandel}
\author[5]{A.~Horneffer}
\author[2]{D.~Huber\corref{cor}}
\ead{Daniel.Huber@kit.edu}
\author[1]{T.~Huege}
\author[1,16]{P.G.~Isar}
\author[10]{K.-H.~Kampert}
\author[2]{D.~Kang}
\author[11]{O.~Kr\"omer}
\author[3]{J.~Kuijpers}
\author[2]{K.~Link}
\author[12]{P.~{\L}uczak}
\author[2]{M.~Ludwig}
\author[1]{H.J.~Mathes}
\author[2]{M.~Melissas}
\author[8]{C.~Morello}
\author[1]{J.~Oehlschl\"ager}
\author[2]{N.~Palmieri}
\author[1]{T.~Pierog}
\author[10]{J.~Rautenberg}
\author[1]{H.~Rebel}
\author[1]{M.~Roth}
\author[11]{C.~R\"uhle}
\author[6]{A.~Saftoiu}
\author[1]{H.~Schieler}
\author[11]{A.~Schmidt}
\author[1]{F.G.~Schr\"oder}
\author[13]{O.~Sima}
\author[6]{G.~Toma}
\author[8]{G.C.~Trinchero}
\author[1]{A.~Weindl}
\author[1]{J.~Wochele}
\author[1]{M.~Wommer}
\author[12]{J.~Zabierowski}
\author[5]{J.A.~Zensus}

\address[1]{Karlsruhe Institute of Technology (KIT), Institut f\"ur Kernphysik, Germany}
\address[2]{Karlsruhe Institute of Technology (KIT), Institut f\"ur Experimentelle Kernphysik, Germany}
\address[3]{Radboud University Nijmegen, Department of Astrophysics, The Netherlands}
\address[4]{Dipartimento di Fisica Generale dell' Universit\`a Torino, Italy}
\address[5]{Max-Planck-Institut f\"ur Radioastronomie Bonn, Germany}
\address[6]{National Institute of Physics and Nuclear Engineering, Bucharest, Romania}
\address[7]{Universit\"at Siegen, Fachbereich Physik, Germany}
\address[8]{INAF Torino, Instituto di Fisica dello Spazio Interplanetario, Italy}
\address[9]{ASTRON, Dwingeloo, The Netherlands}
\address[{10}]{Universit\"at Wuppertal, Fachbereich Physik, Germany}
\address[{11}]{Karlsruhe Institute of Technology (KIT), Institut f\"ur Prozessdatenverarbeitung und Elektronik, Germany}
\address[{12}]{National Centre for Nuclear Research, Department of Cosmic Ray Physics, {\L}\'{o}d\'{z}, Poland}
\address[{13}]{University of Bucharest, Department of Physics, Romania}
\scriptsize{
\address[{14}]{now at: Universidad Michoacana, Morelia, Mexico}
\address[{15}]{now at: Universidad S\~ao Paulo, Inst. de F\'{\i}sica de S\~ao Carlos, Brasil}
\address[{16}]{now at: Institute for Space Sciences, Bucharest, Romania}
}

\cortext[cor]{Corresponding author:}

\begin{abstract}
To better understand the radio signal emitted by extensive air-showers and to further develop the 
radio detection technique of high-energy cosmic rays, the LOPES experiment was reconfigured to LOPES-3D.
LOPES-3D is able to measure all three vectorial components of the electric field of radio 
emission from cosmic ray air showers. The additional measurement of the vertical component ought to increase the reconstruction 
accuracy of primary cosmic ray parameters like direction and energy, provides an improved sensitivity 
to inclined showers, and will help to validate simulation of the emission mechanisms in the atmosphere.
LOPES-3D will evaluate the feasibility of vectorial measurements for large scale applications. 
In order to measure all three electric field components directly, a tailor-made antenna type 
(tripoles) was deployed. 
The change of the antenna type necessitated new pre-amplifiers and an
overall recalibration. 
The reconfiguration and the recalibration procedure are presented and the 
operationality of LOPES-3D is demonstrated.
\end{abstract}


\end{frontmatter}


\section{Introduction}
Cosmic rays have been studied for a long time. Although the detection techniques developed very fast and promisingly, the key questions are still not answered satisfactorily. The process of the acceleration, the origin and the propagation of cosmic rays through outer space are not completely understood. To answer these questions, the cosmic rays that reach the Earth have to be studied with high precision at highest energies. In order to do so, one has to stick to ground based observatories that cover huge areas since the flux of cosmic rays at these energies is too low to measure them directly in space. Instead, extensive air showers \cite{AugerEhrenfestMaze1939} that are caused by high energy cosmic rays have to be observed.
One important parameter when observing cosmic rays indirectly is the shower depth $X_{max}$ which can be used for composition studies. So far fluorescence telescopes reach the best results in the  $X_{max}$ resolution, but the uptime is limited to clear and moonless nights which results in a net uptime of $\leq 15 \%$ \cite{Abraham2010239}. 
Radio detection can provide a very high uptime ($\geq 95$\%) \cite{Haungsradio} and recent simulation studies not taking into account noise have shown, that indeed the radio signal has an intrinsic resolution in $X_{max}$ in the order of $30 \unit{\frac{g}{cm^{2}}} $ \cite{FrankThesis2011}. Therefore the interest of developing  the radio detection technique is large and several approaches have been made to advance and exploit this technique to the maximum. 
Earlier approaches on the radio detection of cosmic rays as done in the 1960's and 1970's  \cite{jelley} suffered from the lack of fast digital electronics available at that time. These first efforts were reviewed by Allan in 1971 \cite{allan}. With the development of digital electronics air shower radio measurements have become feasible.\\
The radio detection technique can be used in hybrid mode with a particle detector air-shower array or as stand alone experiment. To gain maximum information on the radio signal produced by an extensive air shower, it is necessary to measure and to better understand the complete signal. Since the E-field vector of the radio emission is a three dimensional quantity, all three components of this vector need to be detected to not lose any information. A completely measured and reconstructed E-field vector will considerably contribute to proceed in the understanding of the radio emission mechanism during the present development and optimization phase of this new detection technique. With the LOPES-3D setup we want to study the capability of vectorial measurements in air shower observations. LOPES-3D is designed to measure all three components directly and not only a two dimensional projection. Fully understanding the emission mechanism will increase the quality of air shower parameter reconstruction such as $X_{max}$, the energy of the primary particle etc. Polarization studies done by CODALEMA \cite{CODALEMA}, LOPES \cite{ginaicrc2009} and AERA \cite{AERA} have made a huge step in confirming geomagnetic emission which is described with a $\vec{v}\times\vec{B}$\,-dependence\footnote{the dominant emission in air showers is assigned to the geomagnetic effect where the amplitude is in first order proportional to the cross product of the shower axis and the vector of the Earth's magnetic field} as the main emission mechanism in first order, and the time variant charge excess as a second order effect \cite{icrcCODALEMA1,icrcCODALEMA2,icrcAERA}. These measurements were performed with antennas that are sensitive to two components of the electric field vector. Hence with an antenna that is sensitive to three components a significant information gain is expected. With 3D measurements the direction of the E-field vector can be determined with the data obtained with only one station, fixing a plane for the arrival direction. By measuring the vertical component the sensitivity will be increased, in particular for inclined air showers.\\ To be able to detect all three components of the E-field vector directly, the LOPES experiment at the Karlsruhe Institute of Technology (KIT) has been reconfigured. The LOPES experiment is situated within the KASCADE-Grande \cite{kascade,kascade-grande} experiment which provides a trigger and high quality per event air shower information. 
 
\section{Evaluation of possible antennas}
\label{b}
\begin{figure}[h!]
\begin{center}
\includegraphics[width= .4\textwidth ,angle=0]{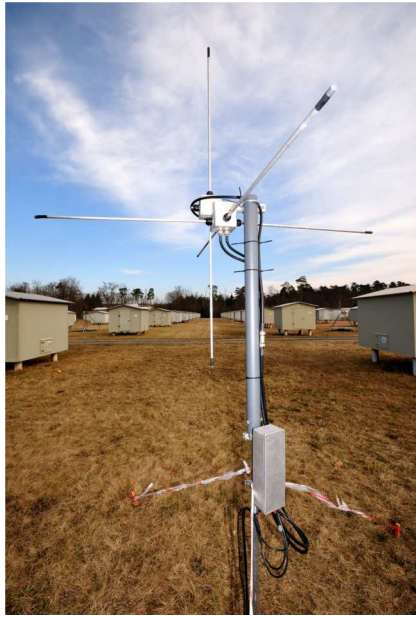} 
\end{center}
\caption{Photography of a tripole antenna station. The metal box at the pole is the housing of the pre-amplifiers.}
\label{tripole}
\end{figure}

For the reconfiguration of the LOPES experiment \cite{FalckeNature2005}  there were two different types of antennas under consideration. The SALLA (Small Aperiodic Loaded Loop Antenna), which is a special type of "Beverage antennas" \cite{KroemerThesis2008}, provides a gain pattern that is hardly sensitive to the ground conditions. This is desirable as it decreases systematic uncertainties. Another advantage, which is not that important for the LOPES setup is that the mechanics of this antenna is very rugged. In the case for LOPES-3D an antenna would consist of two crossed SALLAs to detect the east-west and north-south component of the E-field vector and for the vertical component a dipole would be used. Choosing a different antenna type for the vertical component is necessary since the gain pattern of a SALLA is asymmetric \cite{kroemer}, which will lead to a decrease of sensitivity in a certain direction.\\
The other antenna option was the tripole. Tripole antennas \cite{tripole} are the most straight forward approach to be sensitive to all three electric field components. These antennas are also used in radio astronomy projects like for example LOIS \cite{lois}, the LOFAR \cite{lofar} outrigger in Scandinavia. One tripole consists of three crossed dipoles (c.f. figure \ref{tripole}) and therefore provides a homogeneous setup. The tripole is sensitive to signals that arrive nearly horizontally and offers a generally higher sensitivity compared to the SALLA.\\
The decision to use the tripole was based on test measurements that were performed with both types of antennas. For these measurements three channels of the running LOPES experiment were connected with the SALLA plus dipole and the tripole, respectively. The performance of the tripole was significantly better, i.e. the efficiency in detecting air showers in coincidence with the already running LOPES experiment was higher and the signal-to-noise ratio of artificial beacon\footnote{The beacon emits constant sine signals that are used for the time calibration of LOPES-3D.} \cite{SchroederTimeCalibration2010} (see section \ref{beacon}) signals was distinctly better. During the 4 weeks of testing the tripole, LOPES detected $11$ air showers. All of these $11$ air showers could also be observed with the tripole antenna. In contrast, during 6 weeks of testing only $10$ of $16$ events detected with LOPES could be seen with the SALLA. In addition, it is of high interest to observe a high signal-to-noise ratio (SNR) for the signals emitted by the beacon (table \ref{table2}). For this measurement only 2 channels are of interest since the third channel (vertical) is connected to the same antenna type (dipole) and therefore gives no quality criterion for the antenna type.

\begin{table}[h!]
\centering
\caption{Signal-to-noise ratio of the beacon signals measured with the different prototype antennas over a measurement period of 1 day.}
\begin{tabular}{c|c|c|c}
\multicolumn{4}{r}{Signal-to-noise ratio}\\
\hline
\hline
polarization&frequency [MHz]&SALLA&Tripole\\
\hline
NS & 68.1 & 118 & 185\\
EW & 68.1 & 167 & 397\\
NS & 63.5 & 462 & 793\\
EW & 63.5 & 221 & 915\\
\hline
\end{tabular}
\label{table2}
\end{table}

\begin{figure}[h!]
\begin{center}
\includegraphics[width= .5\textwidth ,angle=0]{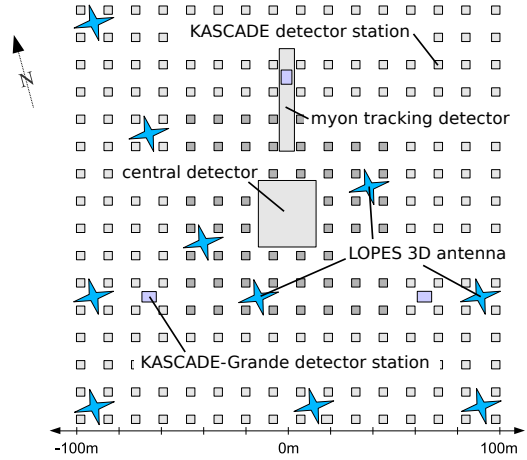} 
\end{center}
\caption{Positions of the LOPES-3D antennas within the KASCADE array. The LOPES-3D antennas are marked as blue stars, the grey squares mark the KASCADE detector huts, where the outer huts consist of unshielded and shielded detectors and the inner huts of unshielded detectors, only. 3 of the 37 KASCADE-Grande stations as well as the muon tracking and the central detector of KASCADE are also shown.}
\label{LOPESantennen}
\end{figure}

Another argument for the tripole is the homogeneous setup which means 
that all three channels are more similar than if using different antenna types.
\\ 
Since the tripole allocates three channels (one channel per direction) and the LOPES experiment provides $30$ channels, the station number was reduced to $10$. With the reconfiguration not only the antenna type and the low noise amplifier (LNA) changed, but also the antenna positions. The positions have to be selected carefully considering different aspects such as
\begin{itemize}
\item covering an area as large as possible for the envisaged primary energy range
\item avoiding regularity to improve the use of LOPES-3D as an interferometer
\item re-using existing cabling and cabling tunnels
\end{itemize}
 It is important to have an irregular grid since LOPES is used as digital radio interferometer, and with a regular antenna grid the side lobes of such an instrument become very pronounced \cite{Almamemo}. The positions of the LOPES-3D antennas within the KASCADE array are shown as blue stars in figure \ref{LOPESantennen}. The LOPES coordinate system is pointing towards the magnetic North whereas KASCADE is orientated with respect to the buildings of the institutes surrounding the array. Both coordinate systems have the same origin but the LOPES system is rotated by $15.23\,^{\circ}$ with respect to the KASCADE coordinate system.

\section{The antenna for LOPES-3D}
\begin{figure}[h!]
\begin{center}
\includegraphics[width= .45\textwidth ,angle=0]{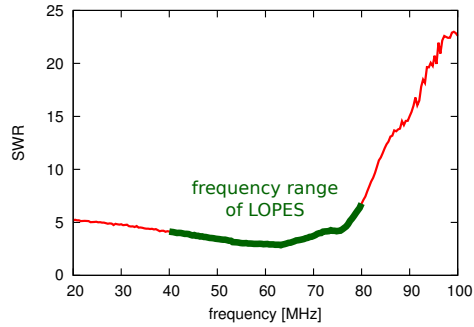} 
\end{center}
\caption{Measured standing wave ratio for one dipole of the tripole. Due to the used filter electronics only the frequency range shown as bold line is of interest.}
\label{swr}
\end{figure}

\begin{figure}[h!]
\begin{center}
\includegraphics[width= .75 \textwidth ,angle=0]{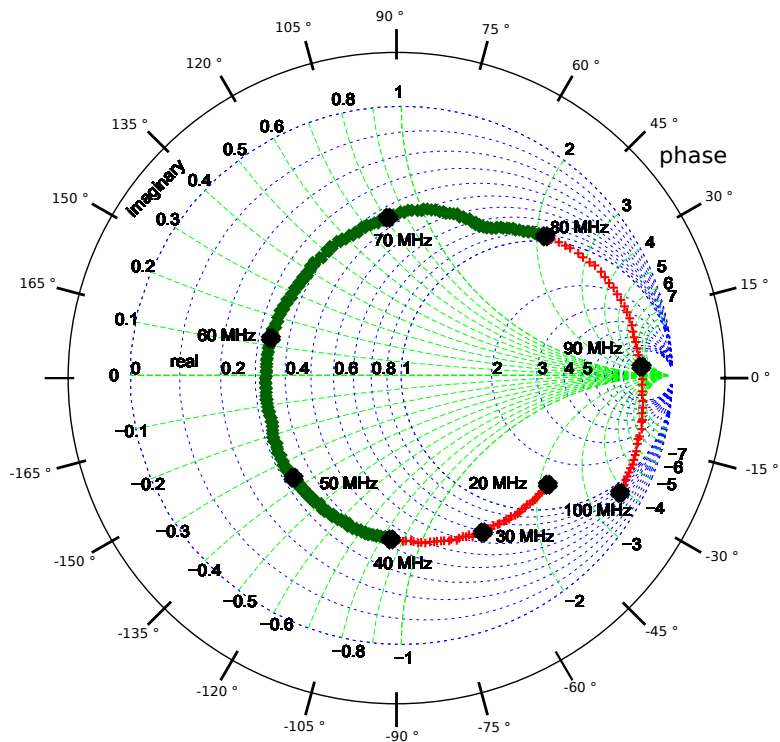} 
\end{center}
\caption{Measured Smith-chart for one dipole of the tripole. The line highlighted in bold green corresponds to the frequency bandwidth of LOPES ($40$ to $80$\,MHz), the red crosses mark the measured range of $20$ to $100$\,MHz.}
\label{smith}
\end{figure}

\begin{figure}[h!]
\begin{center}
\includegraphics[width= .5\textwidth ,angle=0]{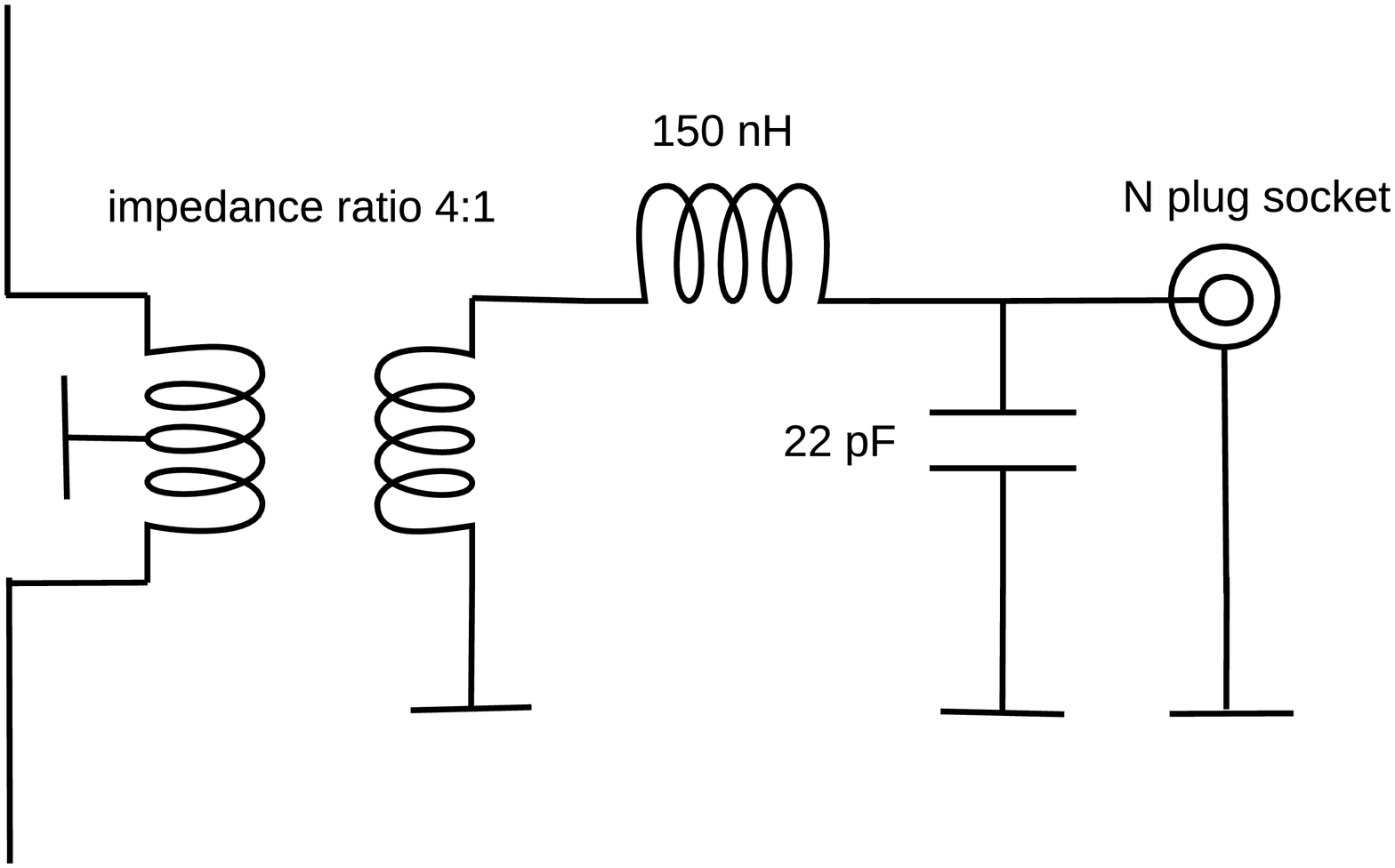} 
\end{center}
\caption{Scheme of the matching of one dipole antenna of the tripole to a channel of the LOPES-3D experiment.}
\label{dipolschema}
\end{figure}

The antenna type used for LOPES-3D is a tripole which consists of three dipoles that are perpendicular to each other. One dipole has a length of $1.3\,$m which is $\frac{\lambda}{4}$ of the central wavelength of the frequency band of LOPES ($40$-$80$\,MHz). Each dipole couples to a coax cable via a Ruthroff balun transformer \cite{balun} including an LC-matching, see figure \ref{dipolschema}. The impedance ratio of $4$:$1$ was chosen to match the characteristic impedance, $\unit{200}{\,\Omega}$, of the antenna. To suppress noise from broadcasting stations the balun is connected with an LC-circuit (low-pass filter). The shielding of the coax cable is grounded. The measured standing wave ratio (SWR) is shown in figure \ref{swr}. Since the dipole is designed as a broadband $\frac{\lambda}{4}$ antenna (length of $1.3$\,m) the SWR has a minimum at the centre frequency of the bandwidth but does not change strongly (standard deviation $0.78$) over the bandwidth of LOPES. It is desirable to have homogeneous antenna characteristics over the whole frequency band. The measured normalized impedance Smith-chart for this antenna is shown in figure \ref{smith}. Here the data are normalized to $\unit{50}{\,\Omega}$. Both, the SWR and the Smith-chart were measured from $20$ to $100$\,MHz, where the interesting region for LOPES ($40$ to $80$\,MHz) is highlighted in bold. From the Smith-chart all the characteristic quantities of an LCR circuit, in this case the dipole antenna, can be derived. For a certain point in the Smith-chart, the SWR can be determined by:
\begin{equation}
\mbox{swr}=\frac{1+\vert r \vert}{1- \vert r \vert}
\end{equation} 
with r the distance between origin and the given point. The impedance and phase can directly be read off. The green lines of the grid mark the imaginary part and the blue lines mark the real part. The phase is the angle between the imaginary axis and the line connecting the origin with the point one wants to characterize.

\subsection{Gain pattern}

\begin{figure}[!h]
\begin{center}
\subfloat[$40$\,MHz]{\includegraphics[width= .45\textwidth ,angle=0]{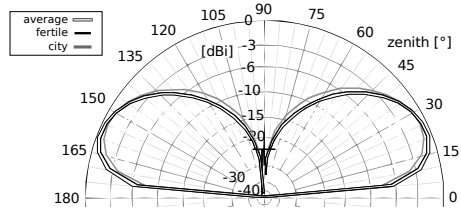}}
  
\subfloat[$60$\,MHz]{\includegraphics[width= .45\textwidth ,angle=0]{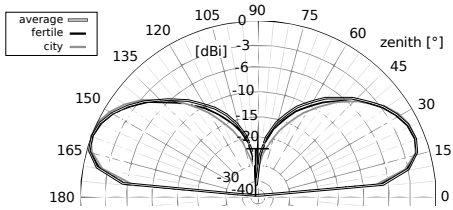}}

\subfloat[$80$\,MHz]{\includegraphics[width= .45\textwidth ,angle=0]{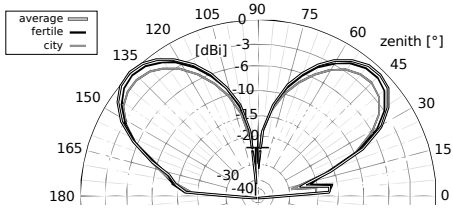}} 
\end{center}
\caption{Simulated gain pattern of the vertical dipole of one tripole at different ground conditions and frequencies. The average ground is used for the LOPES antenna simulation. The different ground conditions are the standard conditions as provided in 4NEC2X \cite{4NEC2}. The dip in the gain pattern for $80$\,MHz originates from the mounting steel pole but does not affect the measurement of cosmic ray air shower radio emission since it only affects the gain pattern at a very insensitive region.}
\label{gainsim}
\end{figure}

The change of the antenna type demanded a new simulation of the gain pattern. Measuring the gain pattern is complicated and requires relatively large facilities to have well defined conditions to ensure the reproducibility of the results. These facilities were not available and we use simulated gain patterns instead. 
For simulating the gain pattern, the Numerical Electromagnetic Code (NEC2) \cite{4NEC2} is used, 
see figure \ref{gainsim}. An average ground was chosen since at the KASCADE site the ground conditions depend on the season, the climate etc. and with the average ground lowest deviations from the different possible conditions are achieved. With these simulations the maximum deviation in the total gain pattern between the different ground conditions is determined to be less than $\unit{0.75}{\,dB}$, including frequency dependence. This of course is a worst case scenario and therefore can be used as maximal systematic error estimation in the simulated gain pattern. The different ground conditions are included in the systematic error since at the LOPES site there is no monitoring for the ground conditions. An even-per-event correction for the ground conditions is therefore not possible. The gain pattern of the horizontally orientated dipoles are in principle the same but rotated by $90^{\,\circ}$. The mounting steel pole has very little influence on the  measurement since it only affects the gain pattern at very insensitive regions. In the analysis the gain pattern is treated in the following way: The gain pattern is simulated in $2$\,MHz frequency steps with a resolution of $5\,^{\circ}$ in azimuth and elevation. This simulation is stored and used for an interpolation which then gives the value needed for the specific analysis.

\begin{figure*}[ht]
\begin{center}
\includegraphics[width= .75\textwidth ,angle=0]{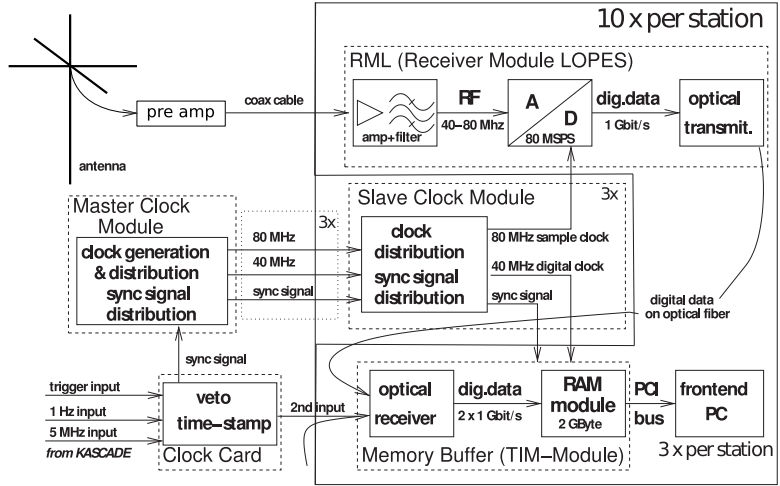} 
\end{center}
\caption{Scheme of the LOPES-3D hardware components adapted from LOPES 30 \cite{NehlsHakenjosArts2007}.}
\label{hardware}
\end{figure*}

\section{Signal chain and data acquisition}
In the following, the hardware components of the LOPES-3D setup are described. 
An overview sketch is shown in figure \ref{hardware}.

\subsection{Pre-amplifiers (LNAs)}
The deployed LNAs are two-channel LNAs with a double bias t-coupling to be fed via phantom feeding with a voltage from $7-24\,$V. At the input of the LNAs there is an over-voltage suppressor and a second order high-pass filter to avoid saturation effects. These LNAs were originally designed for the Auger Engineering Radio Array AERA \cite{Huege2010}. They are based on a  MMIC (monolithic microwave integrated circuit) amplifier module which is unconditionally stable \cite{KroemerThesis2008}.

\subsection{Beacon}
\label{beacon}
\begin{figure}[!h]
\begin{center}
\includegraphics[width= .4\textwidth ,angle=0]{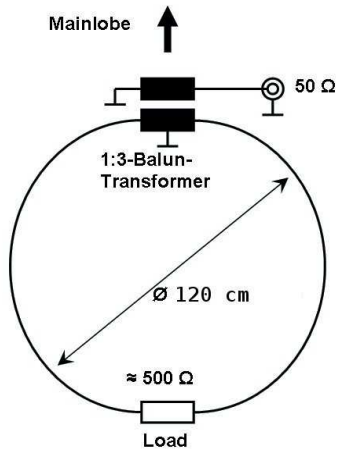} 
\end{center}
\caption{Scheme of the SALLA beacon antenna. The same type was used for the test measurements discussed in section \ref{b}.}
\label{SALLAscheme}
\end{figure}

The beacon is a reference emitter that continuously emits sine waves at three constant frequencies. With the phase differences of these sine waves the timing of the experiment is monitored, improved and corrected for missing sampling cycles \cite{SchroederTimeCalibration2010,FrankThesis2011}. This technique provides an event-per-event calibration of the timing. The beacon was modified to be compatible to the new setup. The beacon antenna type was changed from a dipole antenna to two crossed SALLAs. This was done to test whether the SALLA can be used as beacon antenna for AERA. With two crossed SALLAs that are rotated, it is assured that the signal of the beacon will be seen in all three channels of one station. In addition, the emitting power had to be increased for two reasons: \begin{enumerate}
\item the emitting SALLA has a generally lower gain than the former beacon dipole antenna.
\item the beacon antenna has to be rotated and tilted to be seen in all orientations of the receiving tripole antenna which leads to a loss of power received in the different orientations.
\end{enumerate} A scheme of the SALLA as used for the beacon is shown in figure \ref{SALLAscheme}.

\bigskip
\indent The following hardware components were not changed within the reconfiguration to LOPES-3D but are briefly mentioned here for completeness. A more detailed description is available in ref. \cite{hornefferieee}:

\subsection{Main Amplifier}
The requirements for the main amplifier were $16$\,dB gain, a noise figure
 of less than
$10.3\,\unit{dB}$ \cite{hornphd} and an output intercept point (OIP2) of more than +32\,dBm. This can be
achieved with the commercially available ZFL-500HLN amplifier, which has a noise figure of 3.8 dB, a gain of $19$\,dB, and is therefore far better than the requirements.

\subsection{Filter}
After the transmission through the coaxial cables, the signal is amplified and filtered to a bandwidth of $40$ to $80$\,MHz. This frequency range was chosen since the LOFAR prototype hardware which is used for LOPES has a sampling frequency of 80\,MHz and with this the recordable frequency window is limited to 40 to 80\,MHz when measuring in the 2nd Nyquist domain \cite{Nyquist1928}. In addition, fewest man-made noise is present in this frequency range. The steep edges of the filters result in an effective bandwidth of $43$-$76$\,MHz for ten channels, respectively $43$-$74$\,MHz for channels $11-30$. The different effective bandwidths originate from the different filter modules used for the update from LOPES 10 to LOPES 30. Because of the different effective bandwidths, the data are filtered digitally to a bandwidth of $43$-$74$\,MHz in the analysis software.

\subsection{Digitizer /ADC}
The analog to digital converters sample the signal with a rate of $80$\,MHz
which is provided by a clock distribution board.  When sampling a signal with a frequency which is twice the bandwidth, no information on the signal gets lost since one is operating in the 2nd Nyquist domain \cite{Nyquist1928}. The original signal can be reconstructed by performing an up-sampling. The ADCs have a maximum input voltage of $\pm1$\,V and a resolution of $12$\,bit. 

\subsection{Memory Buffer /TIM boards}
The digitized data is transferred via fibre optics to the memory buffer module. Each module is connected to a PC via PCI-connector and has two inputs. In the $2$\,GB ring buffer either $12.5$\,s when reading one input or $6.25$\,s when reading both inputs can be handled. After an external trigger from \mbox{KASCADE(-Grande)} the data are recorded in total $0.8$\,ms centred around the time of the trigger.

\subsection{Clocks}
To achieve precise timing and a synchronization with KASCADE-Grande, several clock modules have to operate in a synchronized mode. In LOPES, the clocks are synchronized via cables. 

\subsubsection{Clock board}
The clock board receives a $1$\,Hz, a $5$\,MHz and the trigger signal from KASCADE-Grande. This board then generates a sync signal, a time stamp and a veto. The sync signal is transferred to the Master clock module.

\subsubsection{Master clock module}
In this module the $40$\,MHz and $80$\,MHz digital clock are generated and passed through along with the sync signal to the slave clock modules. 

\subsubsection{Slave clock module}
The slave clock modules distribute the $40$\,MHz digital clock and the sync signal to the memory buffer modules and the $80$\,MHz digital clock to the ADCs.

\section{Calibration}
The changed setup of LOPES-3D necessitated a complete recalibration. The individual steps will be explained in the following. A more detailed description of the calibration procedures in general is available in references \cite{NehlsHakenjosArts2007,SchroederTimeCalibration2010}.

\subsection{Dimensions of the antenna grid}
The maximum allowed uncertainty in the position is determined by dividing the maximum allowed uncertainty in time by the speed of light and is in the case of LOPES $\leq 30\,$cm. The positions of the LOPES-3D antennas have been measured with a differential GPS which has an accuracy in the position of $\approx 2\,$cm in x and y and $\approx 2.5\,$cm in z. In addition to this a systematic uncertainty of $5\,$cm has to be taken into account due to the not exactly known position of the phase center of the antenna. With a total uncertainty in the position of $\leq 7.5\,$cm the specification of less than $30\,$cm is clearly fulfilled.

\subsection{Timing}
Within the LOPES experiment, the timing calibration is done in two steps. There is a measurement of the electronics delay as well as an event-per-event calibration done with the beacon. The measurement of the electronics delay is performed in the following way:\\
A pulse generator is connected to the antenna cable instead of the antenna. Both, the pulse generator and the readout are triggered by the regular KASCADE-Grande trigger. The delay between the readout trigger and the emission of the pulse by the pulse generator is always the same and does not need to be known since only the relative timing of the channels is of interest. The time when the pulse appears in the recorded trace is then used to calculate the time shifts between the channels.\\
The beacon signal can be used to monitor and correct timing drifts in the electronics on an event-to-event basis.  Since the phase differences of one sine signal in two channels have always to be constant \cite{FrankThesis2011} they can be used to correct the timing within one period of the sine. Because there are three frequencies available, $53.1$, $65.5$ and $68.1$\,MHz, the timing correction can be performed over more than just one period, and with a higher accuracy. The absolute value of the group delay at these frequencies needs not to be known, because only the differential timing of the different channels is of interest. With the beacon time drifts can be corrected with an accuracy better than $\unit{1}{\,ns}$. In summary with the measurement of the timing with a pulse generator and the monitoring of timing drifts with the beacon, an overall accuracy of $\leq 1$\,ns is achieved.

\subsection{Amplitude calibration}
\begin{figure}[!h]
\begin{center}
\includegraphics[width= .45\textwidth ,angle=0]{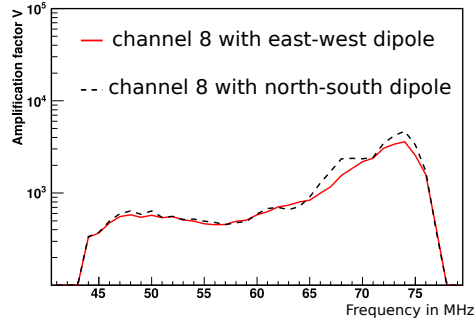} 
\end{center}
\caption{The amplification factors for channel 8 from two measurements. For the first measurement the vertical channel was connected to the north-south oriented dipole, for the second measurement it was connected to the east-west oriented antenna. An overall agreement within $15.6$\% between both measurements can be observed.}
\label{vergleich}
\end{figure}

\begin{figure}[!h]
\begin{center}
\includegraphics[width= .5\textwidth ,angle=0]{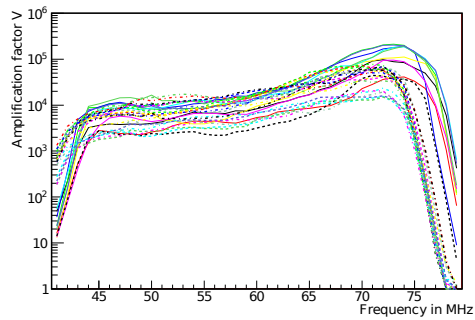} 
\end{center}
\caption{The amplification factors of all 30 channels of the LOPES-3D experiment.}
\label{ampfactor}
\end{figure}

In order to know which field strength at the antenna corresponds to which ADC value the complete electronic signal chain needs to be calibrated. For that purpose, a reference source with a known emission power is arranged above each antenna station to calibrate the individual channels. It is important to:
\begin{enumerate}
\item have a distance between the antenna and the reference source of more than 10 meters, since only at this distances our calibration measurement is valid \cite{NehlsHakenjosArts2007}.
\item measure horizontal position and the height of the reference source by differential GPS with an accuracy of a few cm, which corresponds to an uncertainty in the received power $\leq 2\%$
\item have an alignment of the reference source with the antenna within $\leq 7\,^{\circ}$ deviation which corresponds to $2\%$ variation of the received power due to misalignment of the linear polarizations of the transmitting and receiving antenna.
\item avoid metal parts near the reference source since metal parts can reflect the signal and thereby disturb the calibration measurement.
\end{enumerate}
With the measured ADC values and the calculated field strength at the antenna the frequency dependent amplification factor of each channel can be computed and corrected for in the analysis. \\
The calibration of channels connected to antennas that are vertically orientated is difficult since these antennas are very insensitive to signals from the zenith. Thus a calibration with the reference source above the antenna will suffer large errors from horizontal noise. A calibration with the reference source next to the antenna not high above ground will suffer from reflections from the ground and the KASCADE huts and is therefore not feasible. However, the manufacturing standards of the dipole antennas are very high and the absolute amplitude calibration is performed only for the channel electronics and not the antenna. It is therefore possible to calibrate a channel that was originally connected to a vertically oriented antenna when being connected to a horizontally oriented antenna, see also figure \ref{vergleich}. Hence the best way to calibrate channels connected to vertically orientated antennas is to connect a horizontally orientated antenna instead. \\
The amplification factors in power of the analogue chain (Filter, cable, LNA, main amplifier etc.) of all 30 channels are shown in figure \ref{ampfactor}. These factors describe the frequency dependent attenuation of the signal chain. Features that originate from the different electronics like e.g. steeper filter flanks from the different filters used for $20$ of the 30 channels can be observed.


\section{Monitoring}
\begin{figure}[!h]
\centering
\includegraphics[width= .3\textwidth ,angle=-90]{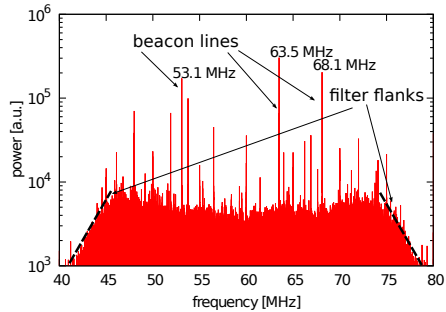}
\caption{Raw spectrum (FFT of the ADCs) taken from the LOPES monitoring. On the Y-axis the power is plotted in arbitrary units and on the X-axis the frequency in MHz. The three beacon frequencies $68.1$, $63.5$ and $53.1$\,MHz can be observed.}
\label{rawspec}
\end{figure}

\begin{figure}[]
\includegraphics[width= .3\textwidth ,angle=-90]{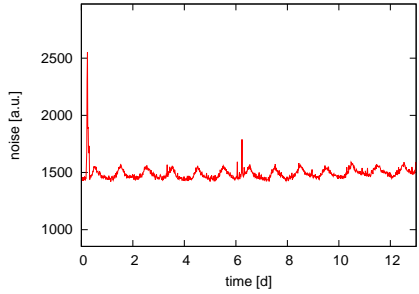}
\caption{The background noise development of channel 24. Starting time noon. A periodic variation can be seen. On the Y-axis the uncalibrated noise is plotted in arbitrary units and on the X-axis the time in hours. The occurring peaks mainly originate from the nearby construction yard.}
\label{krach}
\end{figure}

\begin{figure}[!h]
\subfloat[]{\includegraphics[width= .3\textwidth ,angle=-90]{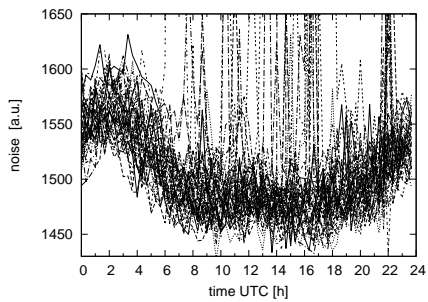}}\\
\subfloat[]{\includegraphics[width= .3\textwidth ,angle=-90]{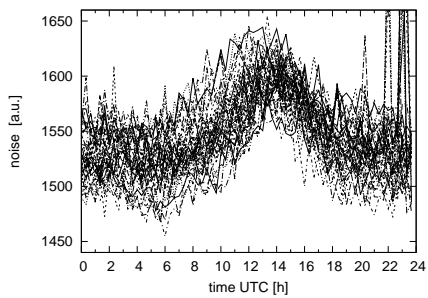}}
\caption{Background noise measured with an east-west aligned antenna over a period of 39 days plotted with a 24 hour period for a measurement from May to June (a) and a measurement from November to December (b) 2011 to check if the noise originates from the galactic transit.}
\label{krach2}
\end{figure}

\begin{figure}[!h]
\includegraphics[height= .3\textwidth ,angle=0]{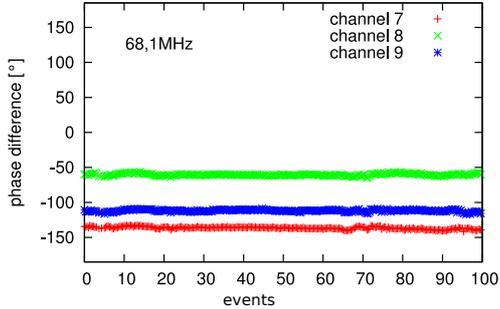}
\caption{The phase differences of the 68.1\,MHz beacon signal of channels 7, 8, 9 in degrees is shown on the Y-axis and on the X-axis the number of events. The measurement of these 100 events corresponds to a time of 33 hours and 20 minutes.}
\label{phase}
\end{figure}

Within the reconfiguration of LOPES, the monitoring was upgraded and improved. Before the upgrade, only the average noise level of each channel was calculated every $20\,\unit{minutes}$ and the vertical atmospheric electrical field was displayed \cite{NehlsThesis2008}.\\ 
During operation of LOPES-3D every $20\,\unit{minutes}$ the last recorded event is analysed in the following way:\\First, an uncalibrated spectrum is derived by performing an FFT (fast Fourier transform) of the raw ADC counts, then the average noise level is calculated, and the most important step is the determination of the present phase differences of the sine signals from the beacon. The uncalibrated spectrum is calculated very fast and gives a good impression of the overall performance of the experiment. An example is shown in figure \ref{rawspec}.\\ 
 These spectra are not corrected for electronic effects such as attenuation in the cables, the gain pattern of the antenna etc. The main purpose of deriving such spectra is to monitor the condition of the experiment. A damaged cable or narrow-band noise sources can be identified very easily without detailed knowledge of the experimental hardware setup. In figure \ref{krach} the average noise level is shown for a period of 13 days. Here a periodic variation can clearly be seen. Such kind of plots are used to monitor the background noise development with time. Moreover deterioration processes of the signal chain can be observed with such plots, e.g., the ageing of an individual LNA will lead to a smaller signal and smaller deviations between maximum and minimum or the stepwise breaking down of a filter module will be seen as a raise in the average noise. It is important to monitor long term changes since it is very hard to identify them on an event-to-event analysis. The background noise can be used to monitor the experiment, when comparing different channels. In figure \ref{krach2} the background noise is shown with a 24 hour period. The drift of the maximum demonstrates that this variation originates from the galactic transit. For the galactic transit a period of $23$ hours and $56$ minutes is expected which corresponds to a drift of $2$ hours per month. A total shift of $12$ hours is expected. At the LOPES site most of the galactic radio emission is buried in anthropogenic noise, therefore the amplitude of the variation is relatively small. With the beacon, LOPES is provided with an event-to-event monitoring of the timing. It is very desirable to monitor this event-to-event time calibration, because this is a very sensitive quantity. For the monitoring the phase differences of the sine waves emitted by the beacon are calculated for every channel and displayed, see figure \ref{phase}.


\begin{figure}[!h]
\begin{center}
\subfloat[east-west]{\includegraphics[width= .3\textwidth ,angle=0]{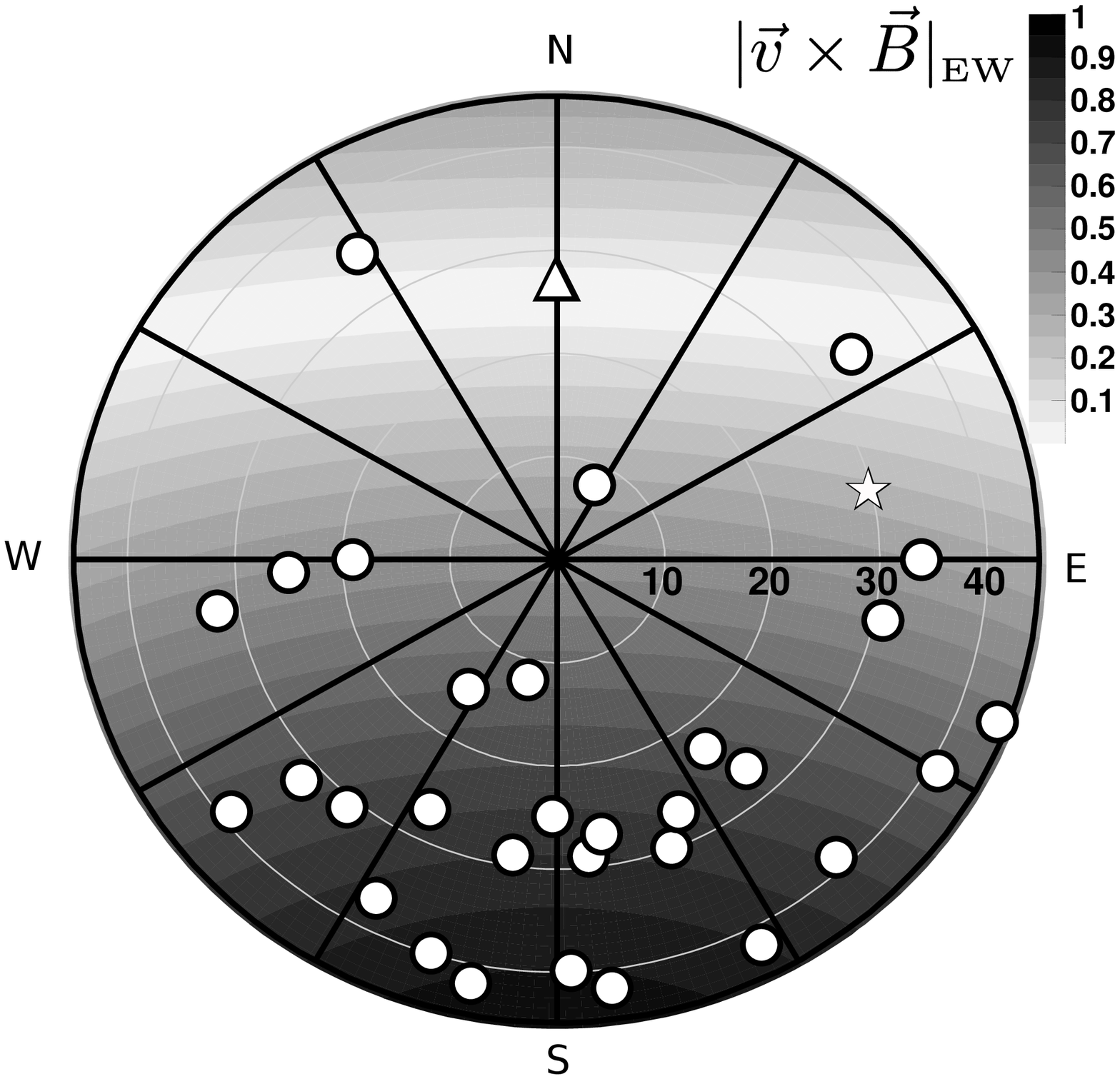}}
  
\subfloat[north-south]{\includegraphics[width= .3\textwidth ,angle=0]{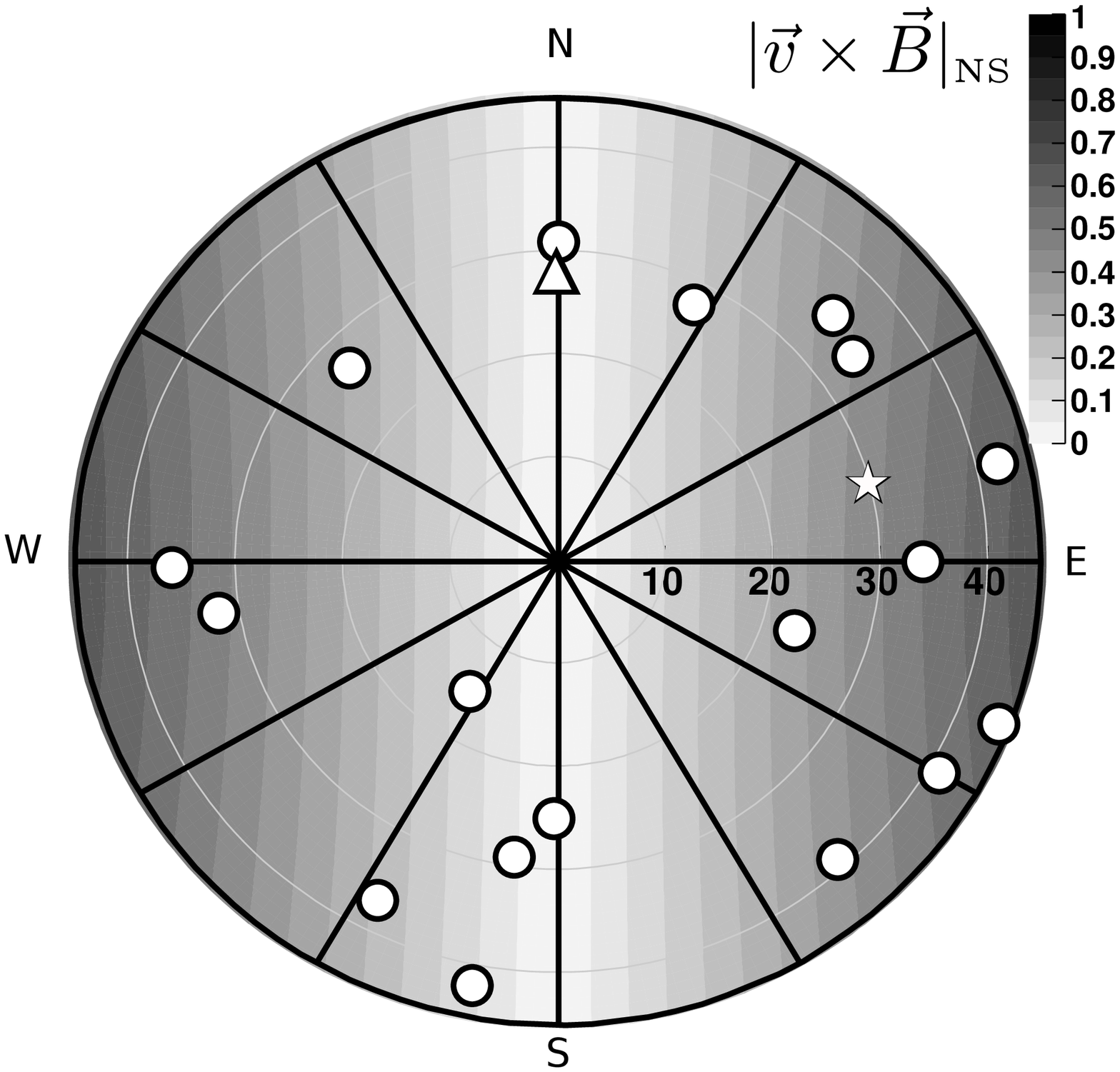}}

\subfloat[vertical]{\includegraphics[width= .3\textwidth ,angle=0]{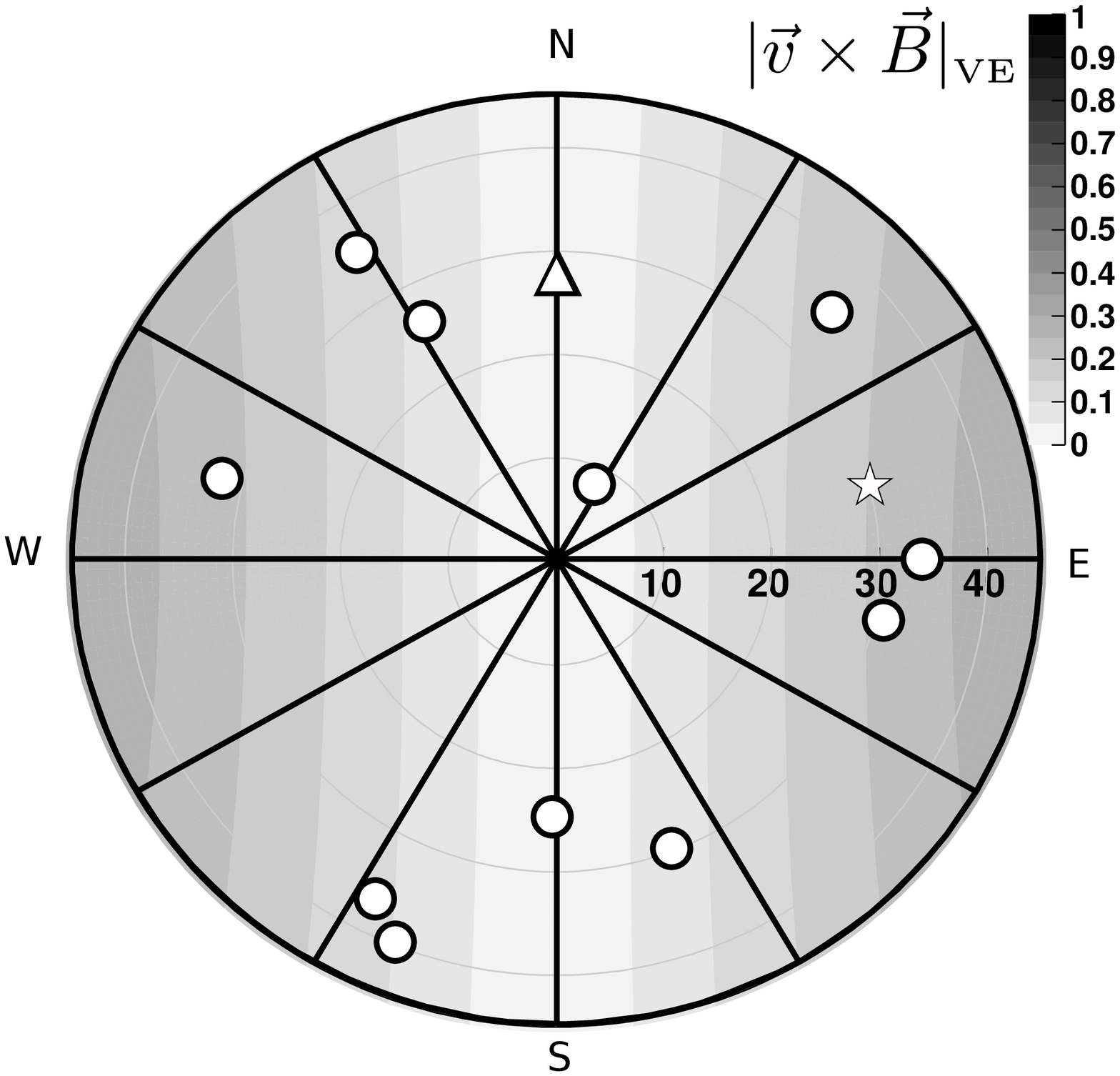}
\label{ve}}
\end{center}
\caption{Arrival directions of measured air showers on top of the expectations of the $\vec{v}\times\vec{B}$\,-Model for the relative amplitudes of the normalized emission vector. The open star marks the event shown in figure \ref{cc} the triangle marks the direction of the magnetic field vector in Karlsruhe.}
\label{vxbcompare}
\end{figure}

\section{First Measurements and Performance}

\begin{figure}[!ht]
\centering
  \subfloat[]{\includegraphics[width=.45 \textwidth]{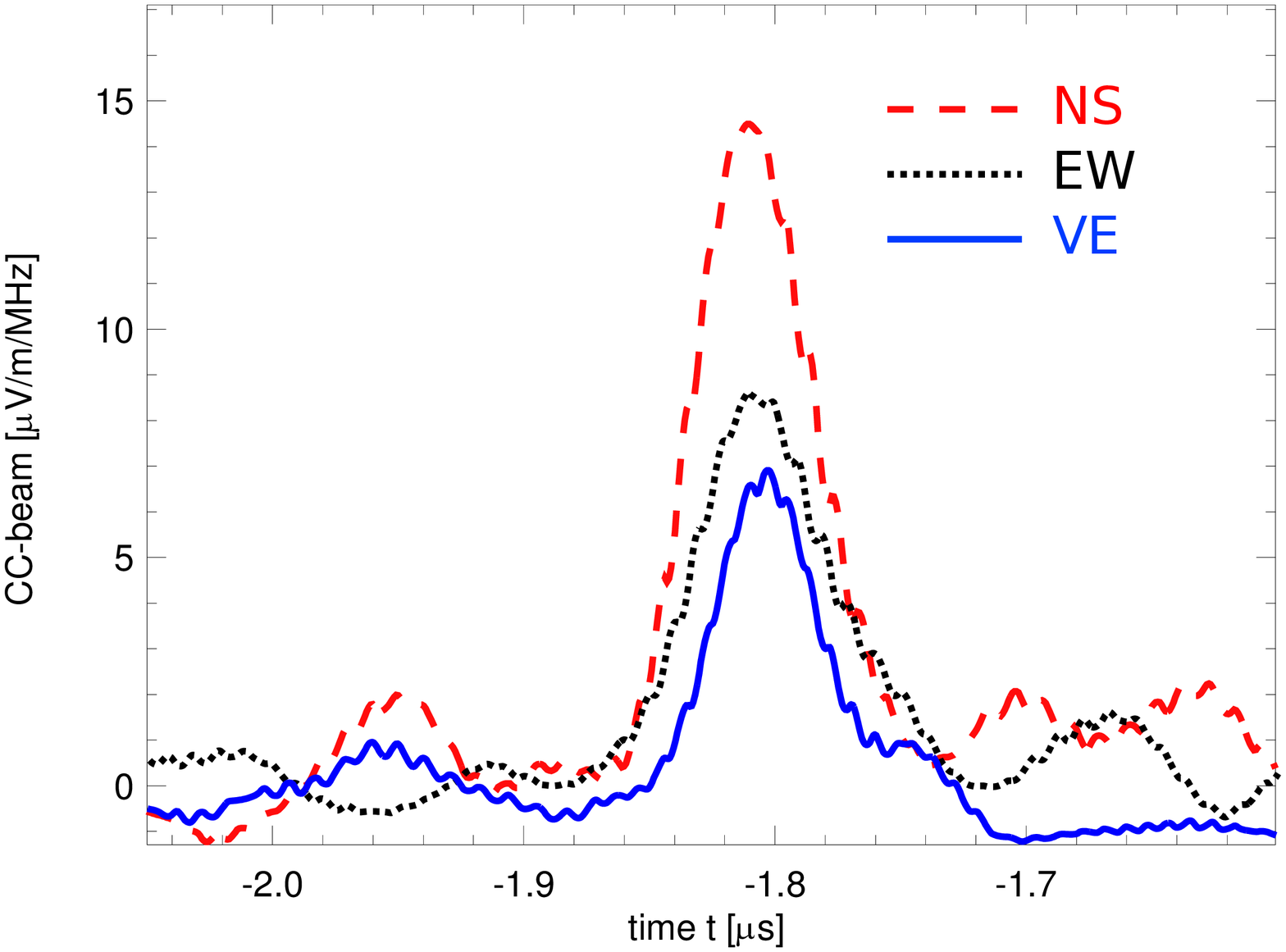}
  \label{figure:5}}\\
  \subfloat[]{\includegraphics[width=.45 \textwidth]{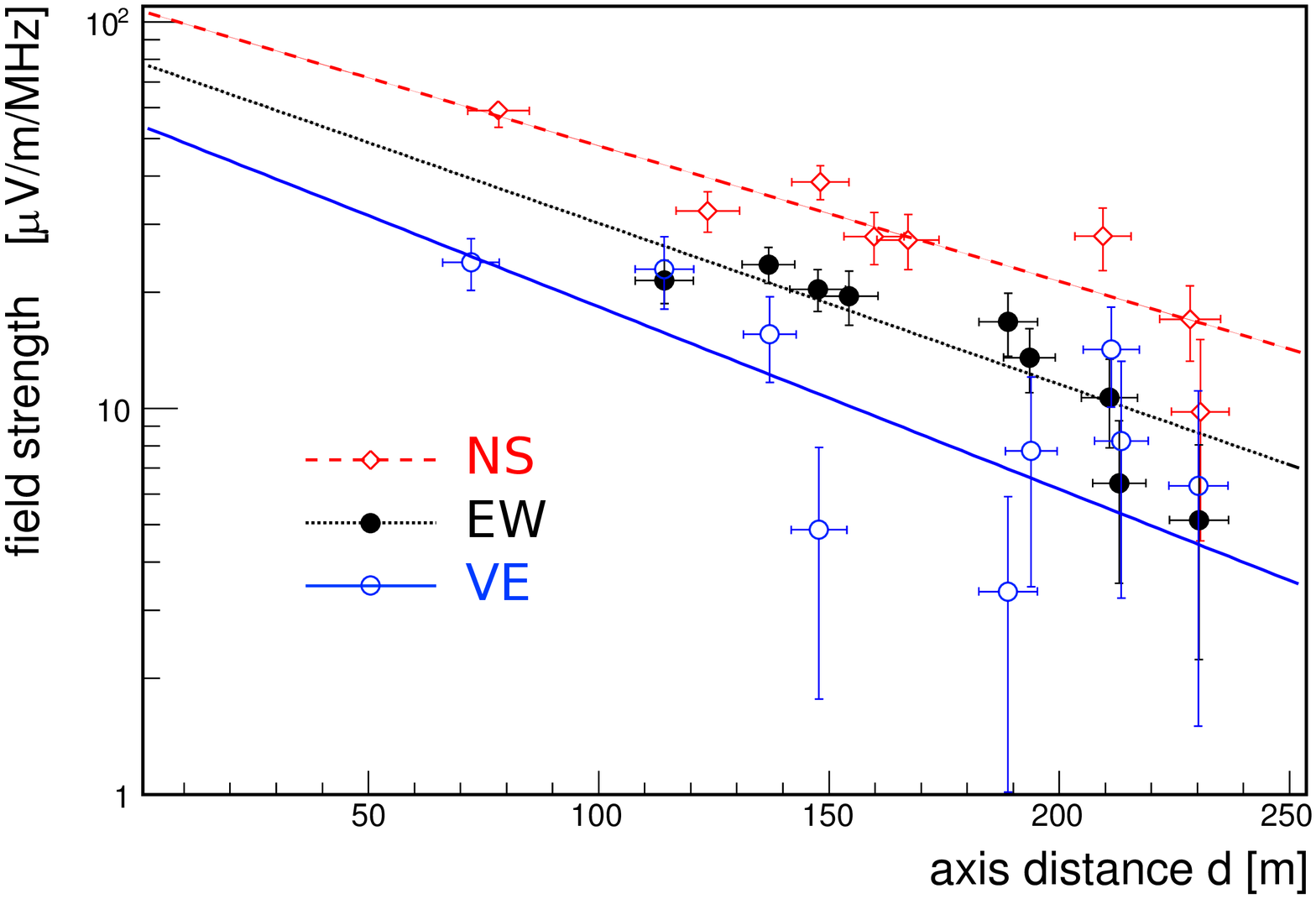}
  \label{figure:5b}}\\
   \caption{Measured CC-beam of an air shower (a) and the according lateral distributions (b). Parameters (from KASCADE-Grande) of this air shower: Primary energy $E_{p}=\unit[8.4\times 10^{17}]{eV}$, Azimuth $\phi=105\,^{\circ}$, Zenith: $\theta=31\,^{\circ}$ }
   \label{cc}
  \end{figure}

In figure \ref{cc} a recorded and reconstructed event in all three vectorial electric field components via digital radio interferometry is shown. The calculated CC-beam from the measured signals of the east-west, north-south and vertically aligned antennas and the corresponding lateral distributions are displayed. For this calculation the beam forming was done for each component separately. The CC-beam is calculated \cite{NiglAA} when using LOPES as digital radio interferometer and gives information on the power that is coherently emitted from a certain direction in the sky:
\begin{equation}
 f{}_{CC}(t)=\sqrt{\frac{1}{N}\sum_{i=1}^{N-1}{\sum_{i>j}^{N}{f_{i}[t]\cdot f_{j}[t]}}}
 \end{equation}
 with $\mbox{N}=\mbox{the number of antennas}$ and $f_{i}[t]=\mbox{time trace of channel i}$.
The air shower shown in figure \ref{cc} had a primary energy of $\unit[8.4\times 10^{17}]{eV}$, arrived with an azimuth angle of $105\,^{\circ}$ and with a zenith angle of $31\,^{\circ}$. For the geomagnetic field in Karlsruhe (inclination\,=\,64\,$^{\circ}$ 45\,$^{\prime}$, declination\,=\,1\,$^{\circ}$ 14\,$^{\prime}$ \cite{magfield}) the $\vec{v}\times\vec{B}$\,-Model predicts for the normalized emission vector $\vec{P}=(0.45|0.81|0.39)$. This is in very good agreement with the measured event which has a normalized emission vector of $\vec{P}=(0.46|0.81|0.36)$. In figure \ref{vxbcompare} the arrival direction of the events reconstructed individually and separately for each vectorial component is plotted on top of the expectations from the $\vec{v}\times\vec{B}$\,-Model. A reasonable agreement is observed which confirms that with LOPES-3D cosmic ray induced air showers can be observed and reconstructed. The event shown in figure \ref{cc} is marked by the star. A detailed combined reconstruction and comparison with expectations for many showers will follow when a reasonable statistic is measured. 

\begin{figure}[!h]
\begin{center}
\includegraphics[width= .5\textwidth ,angle=0]{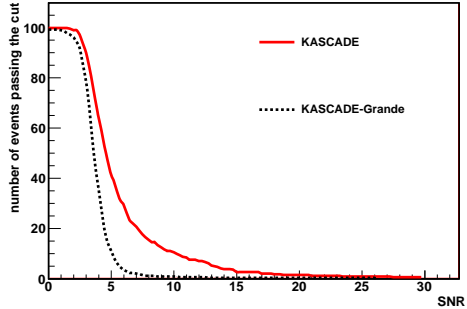} 
\end{center}
\caption{Number of events that pass a certain CC-beam signal-to-noise-ratio cut. KASCADE and KASCADE-Grande means that the core of the air shower is in the respective array and therefore the respective reconstruction is used.}
\label{snrcut}
\end{figure}

For checking the performance of LOPES 3D the signal-to-noise-ratio of the CC-beam is used as a first criterion. The signal height is defined as the maximum height of the Gaussian fit to the CC-beam pulse at $\approx -1.8$\,ns cf. figure \ref{cc}, the noise level as the rms of the CC-beam in the time window of $-204.8$ to $-45.5\,\unit{\mu s}$ before the pulse.
To define a signal-to-noise-ratio (SNR) cut on the CC-beam, the amount of events that pass this cut is plotted over the CC-beam-SNR.
 A drop in the number of events at an SNR value of $5$ for KASCADE triggered and for KASCADE-Grande triggered events can be seen, cf. figure \ref{snrcut}. Triggered means in this context that the shower core is inside the arrays of KASCADE or KASCADE-Grande, respectively.
 The events that have a small SNR value are noise.
After the drop with the increasing SNR only events that have a real air shower signal are counted. With this criterion the statistics shown in table \ref{table1} are obtained.
With LOPES 30 the average rate was $\approx 3.5\,\unit{\frac{events}{week}}$ \citep{NehlsThesis2008} depending on the quality cuts.  
When reducing the number of antenna stations the expected event rate can be calculated. For this estimation we use two different approximations for the detection threshold of a radio interferometer, one which is decreasing linearly when increasing the number of antennas which is a very conservative estimation and second a threshold that goes with the $\sqrt{\mbox{number of antennas}}$. When reducing the antenna positions by a factor of $\frac{1}{3}$, the energy of the primary cosmic ray needs to be 3 ($\sqrt{3}$) times higher to be detected. 
With the index  $\lambda=2$ of the integrated cosmic ray spectrum,
$ f(E) \approx E^{-\lambda} $, the flux of cosmic rays can be estimated to $f={3}^{-2}$ ($f=({\sqrt{3}})^{-2}$) times lower.
Thus the expected event rate of LOPES-3D for a detection threshold that is decreasing linearly when increasing the number of antennas, or a threshold that goes with the $\sqrt{\mbox{number of antennas}}$, respectively is estimated to be a factor of $\frac{1}{9}$ ($\frac{1}{3}$) lower than the one of LOPES 30. 
With $56$ detected events in first 6 months of data taking a detection rate of  $1.75\,\unit{\frac{events}{week}}$ is achieved which is better than the optimistic expectation. This is because LOPES-3D is sensitive to all components of the E-field vector and therefore has a higher sensitivity than a setup with $10$ antennas that are sensitive to only the east-west component of the E-field vector. When only using events that were observed in the east-west direction the event rate of $1.06\,\unit{\frac{events}{week}}$ fits well the expectations.
\begin{table}[!h]
\centering
\caption{Event statistics of LOPES-3D}
\begin{tabular}{ll}
\hline
Average rate & [events/week]\\
\hline
LOPES 30 (EW) & $3.5$\\
LOPES-3D expected (EW) & $0.39-1.17$\\
LOPES-3D (only EW) & $1.06$\\
LOPES-3D (all) & $1.75$\\
\hline
\end{tabular}
\label{table1}
\end{table}

\section{Conclusion}
The LOPES experiment at Karlsruhe Institute of Technology was reconfigured end of 2009 to be now able to measure all three components of the electric field vector from radio emission of cosmic ray induced air showers. The commissioning and testing of the new setup, LOPES-3D, has been successfully completed in May 2010. The experiment is fully operational and performs as expected. Detailed analyses of the data taken will show the prospects of vectorial measurements of the radio emission of cosmic ray induced air showers. The results of the LOPES-3D setup will influence future large scale applications, by evaluating the benefits of vectorial measurements. A fully reconstructed electric field vector allows detailed studies of the different emission mechanisms and thereby may expand the information gain derived by air shower radio observations in the present exploration phase of this new detection technique.

\section*{Acknowledgments}
The authors would like to thank Bo Thid\'e for his helpful suggestions and support before building the LOPES-3D array and the anonymous referees whose remarks led to considerable improvement of the paper.
Part of this research has been supported by grant number VH-NG-413 of the 
Helmholtz Association.
LOPES and KASCADE-Grande have been supported by the German Federal Ministry of Education and Research. KASCADE-Grande is partly supported by the MIUR and INAF of Italy, the Polish Ministry of Science and Higher Education and by the Romanian Authority for Scientific Research UEFISCDI
(PNII-IDEI grant 271/2011).

%

\end{document}